\documentclass[english]{lni}

\usepackage{tabularx}
\usepackage{cite}
\usepackage{comment}
\usepackage[dvipsnames]{xcolor}
\usepackage{graphicx}
\usepackage{url, booktabs}
\usepackage{fancyhdr}
\usepackage{changepage} 
\usepackage{listings} 
\usepackage{tabto}

\usepackage[figurename=Fig., tablename=Tab., small]{caption}

\fancypagestyle{titlepage}{
\fancyhead[RO]{\small F.  Boutros,  N. Damer,  M. Fang,  M. Gomez-Barrero,  K. Raja,  \linebreak C. Rathgeb,   A.  Sequeira,  and M. Todisco (Eds.): BIOSIG 2024, \linebreak Lecture Notes in Informatics (LNI), Gesellschaft f\"ur Informatik, Bonn 2024} 
\fancyfoot{}}

\renewcommand{\headrulewidth}{0.4pt} 
\author{Scott Wellington\footnote{Dept. of Electronic \& Electrical Engineering, University of Bath, United Kingdom, sdlw20@bath.ac.uk} \
Xuechen Liu\footnote{National Institute of Informatics, 2-1-2 Hitotsubashi, Tokyo 101-8430, Japan, \{xuecliu,jyamagis\}@nii.ac.jp} \
Junichi Yamagishi\footnotemark[2]
}

\title{Quantifying Source Speaker Leakage in One-to-One Voice Conversion}

\begin{document}

\maketitle

\renewcommand{\refname}{References}
\setcounter{footnote}{2} 
\thispagestyle{titlepage}
\pagestyle{fancy}
\fancyhead{} 

\fancyhead[RO]{\small Quantifying Source Speaker Leakage in One-to-One Voice Conversion \hspace{25pt}  \hspace{0.05cm}}
\fancyhead[LE]{\hspace{0.05cm}\small  \hspace{25pt} Scott Wellington, Xuechen Liu and Junichi Yamagishi}

\fancyfoot{} 
\renewcommand{\headrulewidth}{0.4pt} 

\begin{abstract}
Using a multi-accented corpus of parallel utterances for use with commercial speech devices, we present a case study to show that it is possible to quantify a degree of confidence about a source speaker’s identity in the case of one-to-one voice conversion. Following voice conversion using a HiFi-GAN vocoder, we compare information leakage for a range speaker characteristics; assuming a `worst-case’ white-box scenario, we quantify our confidence to perform inference and narrow the pool of likely source speakers, reinforcing the regulatory obligation and moral duty that providers of synthetic voices have to ensure the privacy of their speakers’ data.
\end{abstract}


\section{Introduction}

There are always at least three speech data involved in \textit{voice conversion} (VC): the input source speech, the input target speech, and the output converted speech. The most important evaluation criterion for VC is the perceptual similarity between the converted speech and the target speech \cite{tomashenko2022voiceprivacy, cai2023identifying, deng2023catch}.  However, it is essential to evaluate the degree to which the converted speech is also related to the source speech.
This is no different to the objective evaluation of \textit{speaker anonymization}, which concerns the extent to which the converted speech can be tied to the source speaker(s); i.e., how exposed a source speaker is to the risk of a re-identification attack \cite{tomashenko2022voiceprivacy}. 

We must therefore seek to quantitatively evaluate how a source speaker's information is leaking into the converted speech, and develop a VC-adjacent technology to suppress it. In an ideal VC system, the source speech provides only linguistic information. However, particularly with commonly-used Self-Supervised Learning (SSL) architectures such as HuBERT \cite{hsu2021hubert} used in present-day VC applications (e.g. \cite{lin21b_interspeech}), the model encodes a variety of other non-linguistic information which may be used to inform a source speaker's identity.

Two recent attempts have been made to identify the source speaker within VC, and both report that it is possible. One study \cite{cai2023identifying} focused on the deepfake problem used four different VC algorithms, and proposed a system to verify whether the source speakers of converted speech, which sounded like different speakers due to the four VC algorithms, were the same or not. Another forensic study \cite{deng2023catch} proposed a framework for determining whether the VC source speaker is the voice of the suspect by comparing the converted voice with the suspect's natural voice.

The interest of our study is not how to deal with deepfakes and other abuses of VC technologies, but to consider the source speaker's information leakage as a metric for evaluating \textit{privacy-sensitive} VC. This point has not yet been critically considered, but its importance becomes starkly apparent when we think of, for example, personalised synthetic voices for individuals with impaired speech (e.g. \cite{creer2013building, veaux2013towards}). If we consider the case where the output of a text-to-speech (TTS) system, based on the voice of a healthy person (the voice donor), is converted to the voice of a patient with speech disabilities using VC, one can see how the leakage of source speaker information is equivalent to the leakage of training speaker information in the base TTS model. Data privacy is compromised.

The conversion performance of VC is greatly affected by how much the target and source speakers differ in speaker characteristics and accents. When considering the information leakage of the source speaker as an evaluation measure for privacy-sensitive VC, we must therefore not make any assumptions about the use case. Hence, this research uses an SSL-based VC system with a neural speaker encoder to investigate how the source speaker's information leakage changes when VC is performed under both ideal and adverse conditions, including various mismatches of accents and recording environments.

To measure the information leakage, we extract speaker embedding vectors \cite{desplanques2020ecapa} of the converted speech and the source speech, and measure the distribution of their cosine distances. The speaker encoder for this measurement is identical to the speaker encoder utilized in the VC system, hence this is a form a white box testing to analyze the worst-case scenario. \textit{Contra} speaker anonymization, which assumes the use of speaker recognition techniques and standard automatic speaker verification (ASV) metrics, our aim is the assessment of one-to-one VC. Therefore, we use a simple and interpretable measure of Earth Mover's Distance (EMD) to measure the similarity between the above distribution, and the distribution calculated from the source speaker to the target speaker. This metric serves to quantify the source speaker leakage resulting from one-to-one VC.

\section{VC system and measurement of source speaker leakage}

\begin{figure}[t]
\centering
\includegraphics[width=0.8\linewidth]{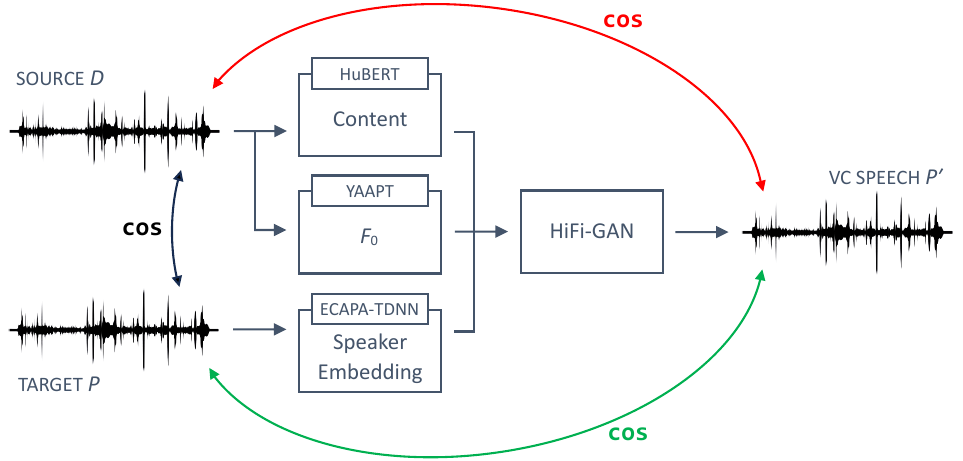}
\caption{An illustration of the pipeline used in this research. Audio from source speaker $D$ and target speaker $P$ are disentangled into speaker embeddings, $F_0$ and speech content for each. The speaker embeddings of source speaker $\textsc{D}$ are replaced with those from target speaker $P$. Remaining disentangled representations are discarded. A HiFi-GAN trained on the LibriSpeech corpus is used to produce the voice-converted (VC) speech $P^\prime$. Cosine similarities are computed between all utterances of $P$, $D$ and $P^\prime$, forming the basis of our calculations for Earth Mover's Distance (EMD) and distributional similarities.} 
\label{fig:methodology}
\end{figure}

Our VC system is similar in design to related studies \cite{van2021comparison, miao22_odyssey}, comprised of content extraction via an SSL model, $F_0$ extraction, a speaker encoder, and a waveform generation module. The content encoder is based on a HuBERT base model  \cite{hsu2021hubert} released by Fairseq  toolkit{\footnote{\label{fairseq}{\url{https://github.com/pytorch/fairseq/}}}} and fine-tuned using \textit{LibriTTS-train-clean-100} \cite{zen2019libritts}. The speaker encoder is based on an ECAPA model \cite{desplanques2020ecapa} trained on the Voxceleb2 dev set \cite{chung2018voxceleb2}, which uses 80-dimensional FBank as inputs. The YAAPT algorithm \cite{kasi2002yet} is used to extract $F_0$. Using the above disentangled features as inputs, the HiFi-GAN vocoder \cite{kong2020hifi} trained using \textit{LibriTTS-train-clean-100} is used to generate waveforms. For one-to-one VC between these two speakers, the disentangled speech content and $F_0$ are extracted from the source speaker and the speaker embedding is extracted from the target speaker to generate converted speech, as shown in Figure \ref{fig:methodology}. 

For the measurement of the source speaker leakage in mismatched VC conditions, we use two data corpora: Speech Databases for Consumer Devices (SPEECON) \cite{iskra-etal-2002-speecon} is a speech database with utterances and recording environments selected to represent realistic data for use with everyday commercial speech devices, and the Voice Cloning Toolkit (VCTK) \cite{VCTK} is a speech database with utterances of read speech in full sentences. Between these two corpora, we cover realistic data for VC in adverse conditions (SPEECON), as well as data recorded in ideal conditions commonly-used to train TTS systems (VCTK). For both, we use all data that comprise speakers from the United Kingdom; this focuses the scope of investigation while still affording analysis of accent mismatch from various regions of the British isles.

SPEECON provides rich metadata for each speaker, comprising gender, accent, age and recording environment. VCTK provides metadata for gender and accent. For each corpus, we first identify the largest suitable subset where these variables are the same: this results in our subset of target speakers. For each experiment, we change one variable: this results in our subset of source speakers. We generate converted speech from these two speaker subsets. For each target and source subset, we calculate the cosine distances between all speaker embeddings: the speaker embedding with the lowest summed cosine distance is identified as the proximal speaker for that subset. As illustrated in Figure \ref{fig:methodology}, we then perform VC between these two proximal speakers. We take the disentangled speech content and $F_0$ from the source proximal speaker ($D$) and the speaker embedding from the target proximal speaker ($P$), to generate converted speech (i.e. this is our proximal voice-converted speaker, $P$$^\prime$), which is expected to be perceptually similar to $P$. 

\begin{figure}[t]
\centering
\includegraphics[width=0.7\columnwidth]{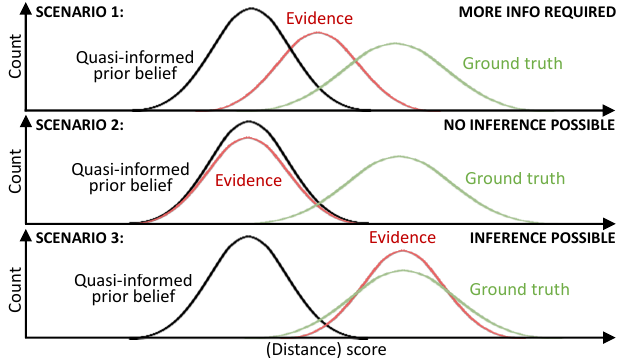}
\caption{Three example scenarios, evaluated through our inference framework. Scenario 1 (top): it is not possible to confidently infer source speaker characteristics from the evidence distribution; information leakage is present, but more data are required to meet the confidence threshold. Scenario 2 (middle): no source speaker characteristics can be inferred; there is no interpretable data leakage. Scenario 3 (bottom): source speaker characteristics can be inferred; there is information leakage.} 
\label{fig:priors}
\end{figure}

We calculate our metrics for distributional comparisons between $P$, $D$ and $P$$^\prime$ to quantify the degree of information leakage that results from such one-to-one VC. The cosine similarity between each speaker is calculated at the utterance level using the same speaker encoder as the VC system \cite{desplanques2020ecapa}, which results in data that are approximately normally distributed. There are many established and robust metrics shared between audio and image processing for security and privatisation of data, of which the Wasserstein distance is interpretable for quantifying the similarity between such distributions. Our pairings of cosine similarities are binned into 50 fixed-width intervals, generating three histograms of distributions: \textsc{cos($P$,$D$)}, \textsc{cos($P$$^\prime$,$D$)} and \textsc{cos($P$$^\prime$,$P$)}. We borrow from the domain of privacy within image processing to calculate the Wasserstein distance between histograms of frequency distributions \cite{xia2015towards}, and use the EMD to quantify the similarity between distributions. EMD is a robust and intuitive metric that expresses the minimum cost of transforming one distribution into another: the higher the Wasserstein distance between two distributions, the higher the EMD.

\begin{figure}[t]
\includegraphics[width=\columnwidth]{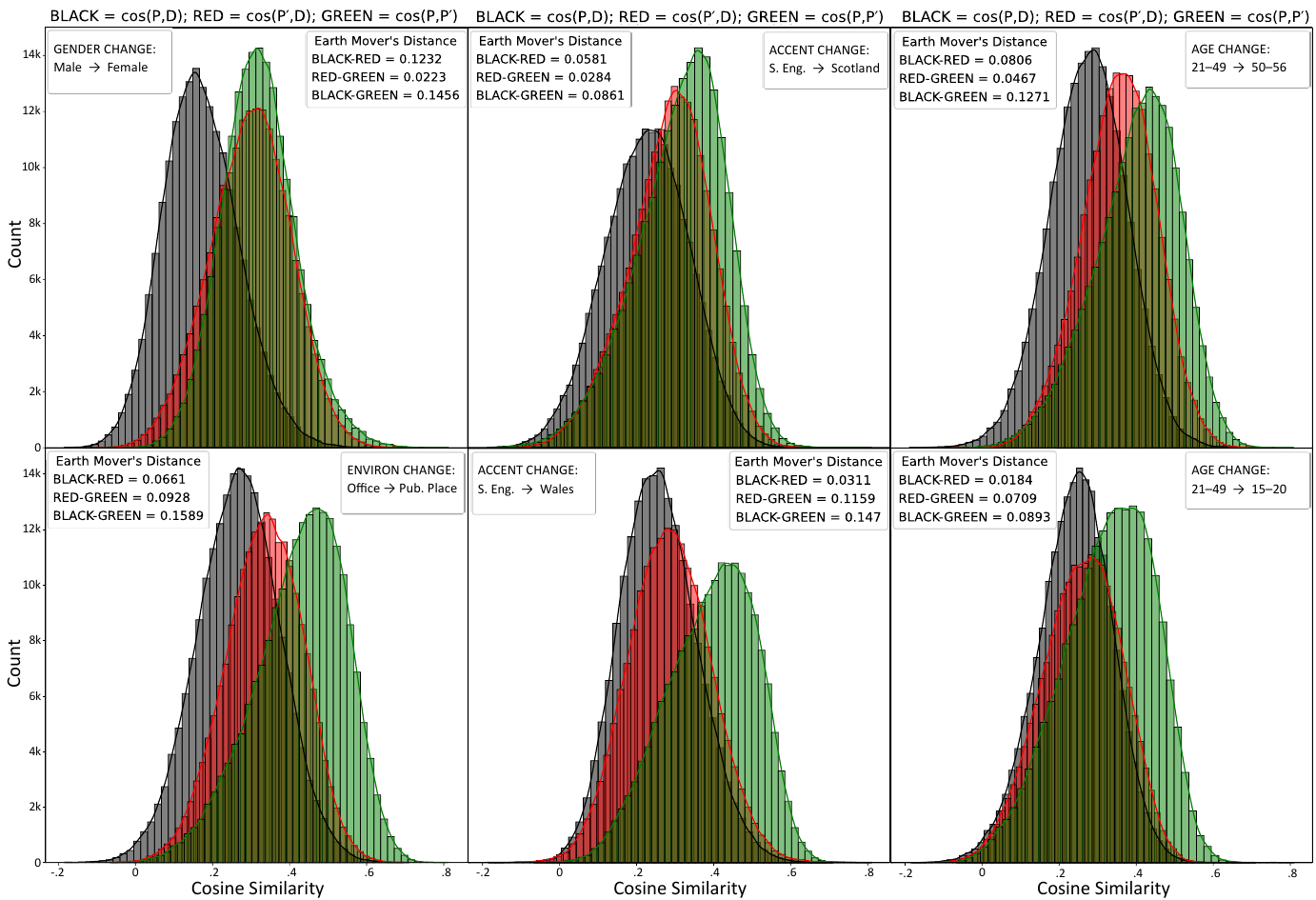}
\caption{Six distributions from SPEECON, with the cosine similarities between the proximal target $P$, proximal source $D$, and the voice converted speech $P^\prime$ (replacing the speaker embeddings of $P$ with those of $D$) binned into 50 fixed-width intervals. The \textsc{\footnotesize{\textcolor{black}{BLACK}}} distributions are \textsc{cos($P$,$D$)}; the \textsc{\footnotesize{\textcolor{red}{RED}}} distributions are \textsc{cos($P^\prime$,$D$)} and the \textsc{\footnotesize{\textcolor{ForestGreen}{GREEN}}} distributions are \textsc{cos($P^\prime$,$P$)}. Plots show the resulting distributional shifts following changes to speaker characteristic variables. Read left-to-right and top-to-bottom, we see \textsc{\footnotesize{\textcolor{red}{RED}}} increasingly move from \textsc{\footnotesize{\textcolor{ForestGreen}{GREEN}}} towards \textsc{\footnotesize{\textcolor{black}{BLACK}}}: a `sliding scale` of how the choice of source speaker characteristics results in greater (or lesser) interpretable data leakage.} 
\label{fig:graphs}
\end{figure}

We follow others \cite{deng2023catch, noe2022towards, noe2023hiding} in assuming terminology and concepts from the well-known Shannon's perfect secrecy \cite{shannon1949communication}, to refer to the `evidence' distribution of cosine similarities, against which we evaluate our `prior belief' distribution of cosine similarities. The farther that our evidence distribution is from our prior belief distribution within the spatial domain, the greater the information leakage from source speaker $D$.

Example scenarios under this inference framework are illustrated in Figure \ref{fig:priors}: when it is impossible to infer attributes from source speaker $D$, given the converted speaker $P$$^\prime$, the distribution \textsc{cos($P$$^\prime$,$D$)} increasingly resembles \textsc{cos($P$,$D$)} as shown in Figure \ref{fig:priors} (middle). Conversely, when it is as easy to guess the attributes of the source speaker $D$ as it is to guess the attributes of the target speaker $P$, given the converted speaker $P$$^\prime$, the distribution \textsc{cos($P$$^\prime$,$D$)} resembles \textsc{cos($P$$^\prime$,$P$)} as shown in Figure \ref{fig:priors} (bottom). Consequently, to use our evaluation metric, the lower the score of \textsc{emd$($cos($P$$^\prime$,$D$),cos($P$$^\prime$,$P$)$)$}, the higher the confidence to infer the attributes of source speaker $D$. We also need to consider the degree to which the attributes of the target speaker $P$ are originally similar to those of the source speaker $D$, which is reflected in \textsc{emd$($cos($P$,$D$),cos($P$$^\prime$,$P$)$)$}.

\section{Results}

\newcommand{\brx}{{\footnotesize EMD}(\textcolor{black}{$B$},\textcolor{red}{$R$})}
\newcommand{\rgx}{{\footnotesize EMD}(\textcolor{red}{$R$},\textcolor{ForestGreen}{$G$})}
\newcommand{\bgx}{{\footnotesize EMD}(\textcolor{black}{$B$},\textcolor{ForestGreen}{$G$})}

\newcommand{\br}{{\footnotesize EMD}(\textcolor{black}{$B$},\textcolor{red}{$R$}) }
\newcommand{\rg}{{\footnotesize EMD}(\textcolor{red}{$R$},\textcolor{ForestGreen}{$G$}) }
\newcommand{\bg}{{\footnotesize EMD}(\textcolor{black}{$B$},\textcolor{ForestGreen}{$G$}) }

\begin{table}[t]
  \caption{EMD of cosine similarities between proximal speakers' utterances, binned into 50 fixed-width intervals, where $P$ is the proximal target speaker, $D$ is the proximal source speaker, $\textcolor{black}{B}$ is \textsc{cos($P$,$D$)}, $\textcolor{red}{\textit{R}}$ is \textsc{cos($P^\prime$,$D$)} and $\textcolor{ForestGreen}{G}$ is \textsc{cos($P^\prime$,$P$)}. Results provided to 4 d.p. for improved interpretability. $n$ is the size of the subset. The speaker characteristics of the proximal target $P$ are: \\\\
  \textbf{SPEECON}: \textit{Gender}= male, \textit{Age} = 21--49, \textit{Accent} = S. England, \textit{Environment} = office \scriptsize($n$ = 30)
  \small\textbf{VCTK}: \quad\quad \textit{Gender} = male, \textit{Accent} = S. England \scriptsize($n$ = 7)} 
  \label{tab:EMDs}
  \centering
  \begin{tabularx}{\columnwidth}{XXXXXX}
    \toprule

    \multicolumn{1}{l}{\textbf{SPEECON}} &
    \multicolumn{1}{l}{\textbf{Prox. source $D$ mismatch}} &
    \multicolumn{1}{r}{\br} &
    \multicolumn{1}{r}{\rg} &
    \multicolumn{1}{r}{\bg} &
    \multicolumn{1}{l}{$L$}
    \\
    \midrule
    
    \multicolumn{1}{l}{\textit{Gender}}
        & \multicolumn{1}{l}{female {\scriptsize ($n$ = 38)}}
        & \multicolumn{1}{r}{$0.1233$} & \multicolumn{1}{r}{$0.0223$} & \multicolumn{1}{r}{$0.1456$} &
        6.5291
        \\
    
    \multicolumn{1}{l}{Age}
        & \multicolumn{1}{l}{15--20 {\scriptsize ($n$ = 9)}}
        & \multicolumn{1}{r}{$0.0184$} & \multicolumn{1}{r}{$0.0709$} & \multicolumn{1}{r}{$0.0893$} &
        1.2595
        \\
    \multicolumn{1}{l}{Age}
        & \multicolumn{1}{l}{50--56 {\scriptsize ($n$ = 9)}}
        & \multicolumn{1}{r}{$0.0806$} & \multicolumn{1}{r}{$0.0467$} & \multicolumn{1}{r}{$0.1271$} &
        2.7216
        \\
    
    \multicolumn{1}{l}{\textit{Accent}}
    & \multicolumn{1}{l}{Midlands {\scriptsize ($n$ = 10)}}
        & \multicolumn{1}{r}{$0.0220$} & \multicolumn{1}{r}{$0.0935$} & \multicolumn{1}{r}{$0.0718$} &
        0.7679
        \\
    \multicolumn{1}{l}{\textit{Accent}}
    & \multicolumn{1}{l}{Wales {\scriptsize ($n$ = 4)}}
        & \multicolumn{1}{r}{$0.0311$} & \multicolumn{1}{r}{$0.1158$} & \multicolumn{1}{r}{$0.1470$}  &
        1.2694
        \\
    \multicolumn{1}{l}{\textit{Accent}}
    & \multicolumn{1}{l}{N. England {\scriptsize ($n$ = 12)}}
        & \multicolumn{1}{r}{$0.0625$} & \multicolumn{1}{r}{$0.0764$} & \multicolumn{1}{r}{$0.1389$}  &
        1.8181
        \\
    \multicolumn{1}{l}{\textit{Accent}}
    & \multicolumn{1}{l}{Scotland {\scriptsize ($n$ = 13)}}
        & \multicolumn{1}{r}{$0.0581$} & \multicolumn{1}{r}{$0.0284$} & \multicolumn{1}{r}{$0.0861$} &
        3.0317
        \\
    \multicolumn{1}{l}{\textit{Accent}}
    & \multicolumn{1}{l}{N. Ireland {\scriptsize ($n$ = 2)}}
        & \multicolumn{1}{r}{$0.0698$} & \multicolumn{1}{r}{$0.1184$}  & \multicolumn{1}{r}{$0.1882$} &
        1.5895
        \\          
    
    \multicolumn{1}{l}{\textit{Environment}}
    & \multicolumn{1}{l}{entertainment {\scriptsize ($n$ = 14)}}
        & \multicolumn{1}{r}{$0.0852$} & \multicolumn{1}{r}{$0.0855$} & \multicolumn{1}{r}{$0.1707$} &
        1.9965
        \\        
    \multicolumn{1}{l}{\textit{Environment}}
    & \multicolumn{1}{l}{car {\scriptsize ($n$ = 22)}}
        & \multicolumn{1}{r}{$0.0872$} & \multicolumn{1}{r}{$0.0806$} & \multicolumn{1}{r}{$0.1677$} &
        2.0806
        \\
    \multicolumn{1}{l}{\textit{Environment}}
    & \multicolumn{1}{l}{public place {\scriptsize ($n$ = 16)}}
        & \multicolumn{1}{r}{$0.0661$} & \multicolumn{1}{r}{$0.0928$} & \multicolumn{1}{r}{$0.1589$} &
        1.7123
        \\
             
    \multicolumn{1}{l}{\textit{Matched to $P$}}
    & \multicolumn{1}{l}{N/A (matched to $P$; {\scriptsize $n$ = 15)}}
        & \multicolumn{1}{r}{$0.0348$} & \multicolumn{1}{r}{$0.0595$} & \multicolumn{1}{r}{$0.0654$} &
        1.0992
        \\        
    \midrule
    \multicolumn{1}{l}{\textbf{VCTK}} &
    \multicolumn{1}{l}{\textbf{Prox. source $D$ mismatch}} &
    \multicolumn{1}{r}{\br} &
    \multicolumn{1}{r}{\rg} &
    \multicolumn{1}{r}{\bg} &
    \multicolumn{1}{l}{$L$}
    \\
    \midrule
    \multicolumn{1}{l}{\textit{Gender}}
        & \multicolumn{1}{l}{female {\scriptsize ($n$ = 10)}}
        & \multicolumn{1}{r}{$0.1028$} & \multicolumn{1}{r}{$0.5882$} & \multicolumn{1}{r}{$0.6902$} &
        1.1734
        \\
    
    \multicolumn{1}{l}{\textit{Accent}}
    & \multicolumn{1}{l}{Midlands {\scriptsize ($n$ = 3)}}
        & \multicolumn{1}{r}{$0.0132$} & \multicolumn{1}{r}{$0.6660$} & \multicolumn{1}{r}{$0.6563$} &
        0.9854
        \\
    \multicolumn{1}{l}{\textit{Accent}}
    & \multicolumn{1}{l}{N. England {\scriptsize ($n$ = 5)}}
        & \multicolumn{1}{r}{$0.0141$} & \multicolumn{1}{r}{$0.7914$} & \multicolumn{1}{r}{$0.6511$} &
        0.8227
        \\
      \multicolumn{1}{l}{\textit{Accent}}
    & \multicolumn{1}{l}{Scotland {\scriptsize ($n$ = 14)}}
        & \multicolumn{1}{r}{$0.0742$} & \multicolumn{1}{r}{$0.6186$} & \multicolumn{1}{r}{$0.6923$} &
        1.1191
        \\
    \multicolumn{1}{l}{\textit{Matched to $P$}}
    & \multicolumn{1}{l}{N/A (matched to $P$; {\scriptsize $n$ = 4)}}
        & \multicolumn{1}{r}{$0.2771$} & \multicolumn{1}{r}{$0.6438$} & \multicolumn{1}{r}{$0.3664$} &
        0.5691
        \\        
    \bottomrule
  \end{tabularx}
  
\end{table}

Table \ref{tab:EMDs} provides the EMD of cosine similarities between speakers' utterances for proximal target $P$, proximal source $D$ and VC-generated $P$$^\prime$, binned into 50 fixed-width intervals. For ease of reading, we will refer to distributions by the colors assigned to them in Figures \ref{fig:methodology}, \ref{fig:priors} and \ref{fig:graphs}: {\footnotesize{\textcolor{black}{BLACK}}}, \textcolor{black}{$B$}, is \textsc{cos}($P$,$D$), {\footnotesize{\textcolor{red}{RED}}}, \textcolor{red}{$R$}, is \textsc{cos}($P$$^\prime$,$D$), and {\footnotesize{\textcolor{ForestGreen}{GREEN}}}, \textcolor{ForestGreen}{$G$}, is \textsc{cos}($P$$^\prime$,$P$).

Note that the difference between \br and \rg should be viewed proportional to \bg, since \textcolor{black}{$B$} and \textcolor{ForestGreen}{$G$} circumscribe the similarity between target speaker $P$ and source speaker $D$. Thus, we may define leakage $L$ as $($\bgx$/$\rgx$)$, with higher values of $L$ indicating greater leakage of source speaker information (Table \ref{tab:EMDs}).

Figure \ref{fig:graphs} illustrates the distributional shift of evidence distribution \textcolor{red}{$R$} from ground truth \textcolor{ForestGreen}{$G$} towards prior belief \textcolor{black}{$B$}. In other words, following one-to-one VC, we are able to rank our mismatched speaker characteristics in terms of the proportional difference of \br and \rg to \bgx: the more that a mismatch results in information leakage, the more that our speaker encoder is able to differentiate between speakers. Here, this `sliding scale' effect can be observed, with more `leaky' mismatched source speaker characteristics (e.g. changing gender; Figure \ref{fig:graphs} top-left) resulting in greater interpretable information leakage, and less leaky mismatches (e.g. changing the age bracket from 21--49 to 15--20; Figure \ref{fig:graphs} bottom-right) resulting in lesser interpretable information leakage.

\section{Discussion}

Our results for intra-group and inter-group VC are similar to the findings of Deng \textit{et al.} \cite{deng2023catch}, who report that their speaker identification system is able to determine the gender of the source speaker (gender$\rightarrow$\textit{same}, or gender$\rightarrow$\textit{different}) with comparatively equal accuracies in either case. With the highest information leakage, gender appears to be a highly-discriminatory speaker characteristic; however, others have noted that inter-gender VC contributes to higher distortion artifacts \cite{tomashenko2022voiceprivacy, ebbers2021contrastive}, so more research is required to determine whether the information leakage is a result of supra-segmental features from the source, or decreased naturalness in the output audio.

In terms of demographics, it notable that for SPEECON and VCTK the speaker characteristic mismatch of \textit{Accent$\rightarrow$Midlands} (and to a lesser extent \textit{Accent$\rightarrow$Wales} for SPEECON) for source speaker $D$ results in low EMD between the evidence and prior belief distributions (when viewed proportionally to EMDs with \textcolor{ForestGreen}{$G$}). In other words, there is little information leakage from source speaker $D$ stemming from these accent characteristics. This result is surprising in the light of SPEECON and VCTK results for \textit{Accent$\rightarrow$Scotland}, where both exhibit higher source speaker information leakage $L$. In other words, information leakage from source speaker $D$ is quantifiably more or less with particular mismatched accents. However, it should be noted that the accent metadata in SPEECON are too imprecise to glean any further insights from these results (we may ask, for instance, what is a `Scotland' accent?), and while the accent metadata in VCTK are more fine-grained (for instance, `Fife, Scotland'), the size $n$ of the subsets may be too small for fair comparison.

For demographic age brackets, we can observe how the confidence to infer source speaker characteristics weakens (Figure \ref{fig:graphs}; top-right and bottom-right) as the age of the proximal source speaker $D$ closer aligns with that of the proximal target speaker $P$ (whose specific age is 26, though embeddings remain proximal within the subset of target speakers, ages 21--49). While differences between inter-group information leakage of demographic characteristics such as age and accent is cursorily mentioned in the literature \cite{srivastava2020evaluating}, there appear to be limited efforts to quantify these differences in a systematic way, and no explorations at all concerning intra-group differences. More research is required in this area.

Our results from changing the environmental variable with SPEECON are interesting: all three indicate a degree of information leakage, which can be counter-intuitive in the light of shared speaker demographic characteristics. Information leakage that stems from identifiable features of different recording environments is often overlooked within VC. Nautsch \textit{et al.} \cite{nautsch2019preserving} note in passing that environmental voice features, as well as a speaker's biological features, are a consideration for unlinkability efforts, but this is not the focus of their study and not considered further. Noe \textit{et al.} \cite{noe2020adversarial} explore information leakage using an adversarial autoencoder, reporting that the disentanglement of speaker characteristics can be reliably used to determine gender; however, although they note that noise information disentanglement (from different recording environments) may also be used to identify speakers in a similar manner, this also not the focus of their study and not explored further.

Bäckström \cite{backstrom2024privacy} notes not only that different recording environments can encode identifying information (speaker linkability based on ambient noise, as well as reverberation and additive background voices), but that attempts to isolate the speaker's voice from the environment can introduce extra signal information that may counter-productively provide additional insights concerning the the source speaker. Our results reinforce this observation that environmental voice features can encode information that increases source speaker linkability. As previously noted, SSL features encompass not only textual information but also various other features; when disentanglement necessarily carries environmental noise from the source speaker's audio, informative signal features from mismatched recording environments may supplement conclusions drawn from inference performed using speaker demographic characteristics.

Our results should be additionally viewed through the lens of model mismatch: the HiFi-GAN vocoder \cite{kong2020hifi} used in this investigation was trained using the \textit{LibriTTS-train-clean-100} subset of the LibriSpeech corpus, which comprises US speakers. When the model is less able to differentiate between speakers (lower scores for \brx) this may be an artefact of the model; in other words, the model's training data is the inhibiting factor, rather than less information leakage stemming from less `leaky' source speaker characteristics. A similar observation is also noted by Cai \textit{et al.} \cite{cai2023identifying}, who also note that, for GAN-based models, the `cycle consistency mechanism' may constrain the generated output in ways that retain speaker-irrelevant features; scores for \br may therefore be biased lower as a result, rather than as a consequence of less information leakage. We acknowledge that we have used a single HiFi-GAN model in this work as a case study: future research with other variations of HiFi-GANs will illuminate further the role of these models in contributing to source speaker leakage.

We note that such artefacts originating from the VC process motivated our use of the `proximal' speaker within candidate speaker pools. Preliminary investigations to create a `representative' source speaker by calculating an average speaker embedding (and using the $F_0$ and content from the proximal target speaker) introduced artefacts into the generated $P$$^\prime$ samples that resulted in our distributions for \textcolor{red}{$R$} exhibiting lower cosine similarity than those of \textcolor{black}{$B$}. Despite these observations, Deng \textit{et al.} \cite{deng2023catch} note that, while high-fidelity VC systems feature less information leakage from the source speaker, there regardless exist no techniques presently to disentangle linguistic, supra-segmental and segmental data from audio that perfectly isolate all features without information leakage.

A similar effect was also reported by Cai \textit{et al.} \cite{cai2023identifying}, who note that speaker identification can be improved by having utterances (here, those of $P$$^\prime$) generated from multiple vocoders, as this helps to better generalize outside the bias or artefacts introduced from a single model. A natural point of future investigation is to therefore examine the extent to which \br is sensitive to information leakage originating from the model itself, since there may be interaction effects between the choice of model to generate $P$$^\prime$, and the speaker demographic and environmental characteristics explored here.

\section{Conclusion}

In this work, we have presented a case study to show that it is possible to quantify a degree of confidence in identifying a source speaker in the case of one-to-one voice conversion, using an interpretable measure of Earth Mover's Distance. We find this is possible due to information leakage, despite simple one-to-one mapping used between source and target. The identity of the source speaker can therefore be compromised. Since providers of synthetic voices must fulfil legal and moral obligations to protect the identities of their source speakers, methodologies to dampen information leakage or obfuscate those identifying features must be pursued.

\bibliography{lniguide}

\end{document}